\documentclass[12pt]{article}
\pdfoutput=1
\usepackage{amssymb}

\usepackage[totalheight = 23cm, totalwidth = 17cm]{geometry}
\usepackage{amssymb,amsmath,amsfonts,amsbsy,graphicx}

\newcommand{\mpl}{m_{\rm Pl}}
\newcommand{\rh}{{\rm rh}}
\newcommand{\calO}{{\cal O}}
\newcommand{\calP}{{\cal P}}
\newcommand{\calR}{{\cal R}}

\begin{document}

\begin{titlepage}

\rightline{\footnotesize{APCTP-Pre2015-001}}

\begin{center}

\vskip 1.0cm

\LARGE{\bf Probing reheating with primordial spectrum}

\vskip 1.0cm

\large{
Jinn-Ouk Gong$^{a,b}$,
\hspace{0.2cm}
Godfrey Leung$^{a}$
\hspace{0.2cm} and \hspace{0.2cm}
Shi Pi$^{a}$
}

\vskip 0.5cm

\small{\it
$^{a}$Asia Pacific Center for Theoretical Physics, Pohang 790-784, Korea
\\
$^{b}$Department of Physics, Postech, Pohang 790-784, Korea
}

\vskip 1.2cm

\end{center}

\begin{abstract}

We study the impacts of reheating temperature on the inflationary predictions of the spectral index and tensor-to-scalar ratio. Assuming sinusoidal oscillations and that reheating process is very fast, the reheating temperature can be constrained for sinusoidal oscillation within a factor of 10 - 100 or even better with the prospect of future observations. Beyond this, we find that the predictions can also be insensitive to the reheating temperature in certain models, including Higgs inflation.

\end{abstract}

\end{titlepage}

\setcounter{page}{0}
\newpage
\setcounter{page}{1}

\section{Introduction}

Recent observational advances place more stringent bounds on the predictions of inflation, for example the Planck mission~\cite{Ade:2015lrj}. When the modes relevant for the cosmic microwave background (CMB) observations exit the horizon during inflation, the slow-roll approximation is supposed to be working very well. Consequently, the properties of the primordial curvature perturbation, which after inflation becomes the seed of the observed temperature fluctuations in the CMB as well as large scale distribution of galaxies, are determined by the form of the inflaton potential in simple canonical models~\cite{Lyth:2009zz}. For example, the spectral index $n_\calR$ and the tensor-to-scalar ratio $r$, see \eqref{index} and \eqref{ratio} respectively. More precise estimates of them, however, require full understanding of the thermal history of the universe beyond the end of the inflationary phase. The largest terra incognita is the stage between the end of inflation and the beginning of the standard radiation dominated epoch, during which the energy in the inflaton sector is transferred to the standard model sector~\cite{reheatingreview}. The detail of this reheating process is much less understood, as the huge range the reheating temperature $T_\rh$ may lie within indicates, 1 MeV $\lesssim T_\rh \lesssim 10^{16}~\text{GeV}$\footnote{The upper bound may be lowered to $10^9$ GeV in order to avoid gravitino overproduction if one considers supersymmetric models~\cite{gravitino}.}.

Nevertheless, like single field slow-roll inflation, there is a simple, minimal scenario of reheating~\cite{reheating}. In this scenario, the inflaton oscillates around the minimum of its effective potential after inflation ends, with a decreasing amplitude due to the damping by the cosmic expansion, until the expansion rate $H$ falls down below the decay rate $\Gamma$ of the inflaton. Then the energy of the inflaton sector is transferred into the standard model sector efficiently and the universe becomes dominated by relativistic particles, corresponding to a reheating temperature $T_\rh \sim \sqrt{\mpl\Gamma}$~\cite{Kolb:1990vq}. Therefore in this minimal scenario of reheating, there are two parameters needed for deriving the duration of reheating, $N_\text{rh}$: an effective equation of state $w_\rh$ to describe the expansion rate during the oscillating phase of the inflaton, and the decay rate $\Gamma$ to determine when reheating happens. The constraint to the reheating temperature can be derived from the observational constraints of $n_\mathcal{R}$ and $r$ once $w_\rh$ is known. Given the potential form, $w_\rh$ can be derived, for instance as in the case of power-law chaotic inflation~\cite{Turner:1983he}.  For a typical case where the local minimum is quadratic and $w_\rh \approx 0$, in a universal class of model where $n_\calR$ is inversely proportional to the number of $e$-folds during inflation, we can constrain $T_\rh$ within a factor of $100$ or so. With future precision experiments, this constraint can be further reduced to a factor of $\calO(1)$ or even better.

If the effective potential is not quadratic at the minimum where $w_\rh\neq 0$, the above estimation for the reheating temperature may not be valid. In an extreme case where $w_\rh=1/3$, the observational data cannot give any constraint on $T_\rh$. Equivalently, it is secure to say that the predictions for $n_\calR$ and $r$ become robust and independent of the detail of reheating in this case if we assume the minimal scenario of reheating. We find that the $\phi^4$ model, even with a non-minimal coupling to gravity such as the case of Higgs inflation, has this property.

This article is organized as follows. In Section~\ref{sec:general}, we revisit how the post-inflationary evolution, particularly reheating, leads to a theoretical uncertainty in the number of $e$-folds of expansion during inflation, $\Delta N_k$. This theoretical uncertainty is then translated into those of the inflationary model predictions such as $n_\calR$ and $r$, which depend on the effective equation of state of the inflaton during reheating in the minimal scenario. We illustrate this using chaotic inflation with a monomial potential as an example as in~\cite{monomial,Martin:2014vha}. We then move on discuss how one might use this theoretical uncertainty due to reheating to constrain the reheating temperature in some general universality classes of models in future precision experiments. In Section~\ref{sec:nonminimal}, we then discuss a specific example, the non-minimally coupled $\phi^4$ model, where the model predictions are robust and independent of the detail of reheating in the minimal scenario of reheating. Section~\ref{sec:conc} is devoted to a short summary.

\section{Reheating corrections in single field inflation}
\label{sec:general}

We consider a certain mode of the comoving wavenumber $k$ which exits the horizon during inflation at $k = a_kH_k$. Considering the ratio of this scale to the size of the present horizon $a_0H_0$, we can relate it to various scales that correspond to critical moments in the cosmic history as~\cite{Liddle:2003as}
\begin{equation}\label{history}
\log \left( \frac{k}{a_0H_0} \right) = \log \left( \frac{a_k}{a_e} \frac{a_e}{a_\rh} \frac{a_\rh}{a_0} \frac{H_k}{H_0} \right) = -N_k - N_\rh + \log \left( \frac{a_\rh}{a_0} \right) + \log \left( \frac{H_k}{H_0} \right) \, ,
\end{equation}
where the subscript $e$ and rh denote, respectively, the end of inflation and the onset of radiation dominated era, viz. the moment of reheating. $N_k\equiv\log(a_e/a_k)$ and $N_\rh\equiv\log(a_\rh/a_e)$ are the $e$-folding numbers between the end of inflation and the horizon crossing of the mode of our interest, and the moment of reheating, respectively. Here $N_\rh$ is highly dependent on the detail of reheating and is related to the reheating temperature $T_\rh$ which appears implicitly in $a_\rh/a_0$. Assuming the conservation of entropy after reheating, $a_\rh/a_0$ can be written as~\cite{Kolb:1990vq}
\begin{equation}\label{Trh}
\frac{a_\rh}{a_0} = \left( \frac{11}{43}g_{s*} \right)^{-1/3} \frac{T_0}{T_\rh} \, ,
\end{equation}
where $g_{s*}$ is the effective number of light species for entropy at the moment of reheating.

Thus without detailed knowledge of the reheating process, any attempt to estimate $N_k$ in \eqref{history} is fundamentally restrictive even after we specify a model of inflation. In the following, we assume the minimal reheating scenario~\cite{reheating} where after inflation ends the inflaton coherently oscillates around the minimum of the potential, followed by decay to relativistic particles. If the decay is fast enough, $N_\rh$ then depends only on how the inflaton oscillates around the minimum and is in principle also determined once a model of inflation is given. It can be written as~\cite{monomial}
\begin{equation}
N_\rh = \frac{1}{3(1+w_\rh)} \log\left( \frac{\rho_e}{\rho_\rh} \right) \, .
\end{equation}
Here, let us define $\beta \equiv \rho_\rh/\rho_e \leq 1$. This means that the reheating temperature can be written as
\begin{equation}\label{Trh}
T_\rh = \left( \frac{30}{\pi^2g_*}\beta\rho_e \right)^{1/4} \, ,
\end{equation}
with $g_*$ being the effective number of relativistic species at the moment of reheating, that is not necessarily the same as $g_{s*}$.

Then, from \eqref{history} $N_k$ is estimated by
\begin{align}\label{Nk}
N_k & = -\log\left( \frac{k}{a_0H_0} \right) + \log\left( \frac{T_0}{H_0} \right) +  \log\left( \frac{H_k}{\rho_e^{1/4}} \right) + \log \left[ \left( \frac{11}{43}g_{s*} \right)^{-1/3} \left( \frac{\pi^2}{30}g_* \right)^{1/4} \right]
\nonumber\\
& \quad + \frac{1-3w_\rh}{12(1+w_\rh)}\log\beta \, .
\end{align}
For a fixed comoving scale $k$, the first and second terms on the right hand side can be very accurately determined by observations, whereas the third term and fourth term depend respectively upon the model of inflation and the underlying particle physics models after reheating ends. For realistic particle physics models, $g_{s*} \sim g_{*} = \calO(100)$. Given $N_k$ only depends on $g_{s*}$ and $g_{*}$ logarithmically, it is accurate enough to take the general case where $g_{s*} = g_{*} = \calO(100)$. Together they give the upper limit of $N_k$ as long as $w_\rh\leq 1/3$. In general these contributions depend only weakly on the detail of the inflation model and the thermal history after reheating, as the dependence is logarithmic. The last term represents the correction from reheating. Once we find $N_k$, or $\phi_k$ precisely, we can accurately evaluate the spectral index and tensor-to-scalar ratio, including higher-order slow-roll corrections~\cite{higherSR}
\begin{align}
\label{index}
n_\calR-1 & = -6\epsilon + 2\eta + \left( 24\alpha-\frac{10}{3} \right)\epsilon^2 - (16\alpha-2)\epsilon\eta + \frac{2}{3}\eta^2 + \left( 4\alpha+\frac{4}{3} \right)\xi^2 \, ,
\\
\label{ratio}
r & = 16\epsilon \left[ 1 + \left( -4\alpha+\frac{11}{3} \right)\epsilon + \left( -2\alpha+\frac{5}{3} \right)\eta \right] \, ,
\end{align}
where $\epsilon \equiv \mpl^2(V'/V)^2/2$, $\eta \equiv \mpl^2V''/V$, $\xi^2 \equiv \mpl^4V'V'''/V^2$ and  $\alpha \equiv 2-\log2-\gamma \approx 0.729637$ with $\gamma \approx 0.577216$ being the Euler-Mascheroni constant.

\begin{figure}[h]
 \begin{center}
  \includegraphics[width=0.65\textwidth]{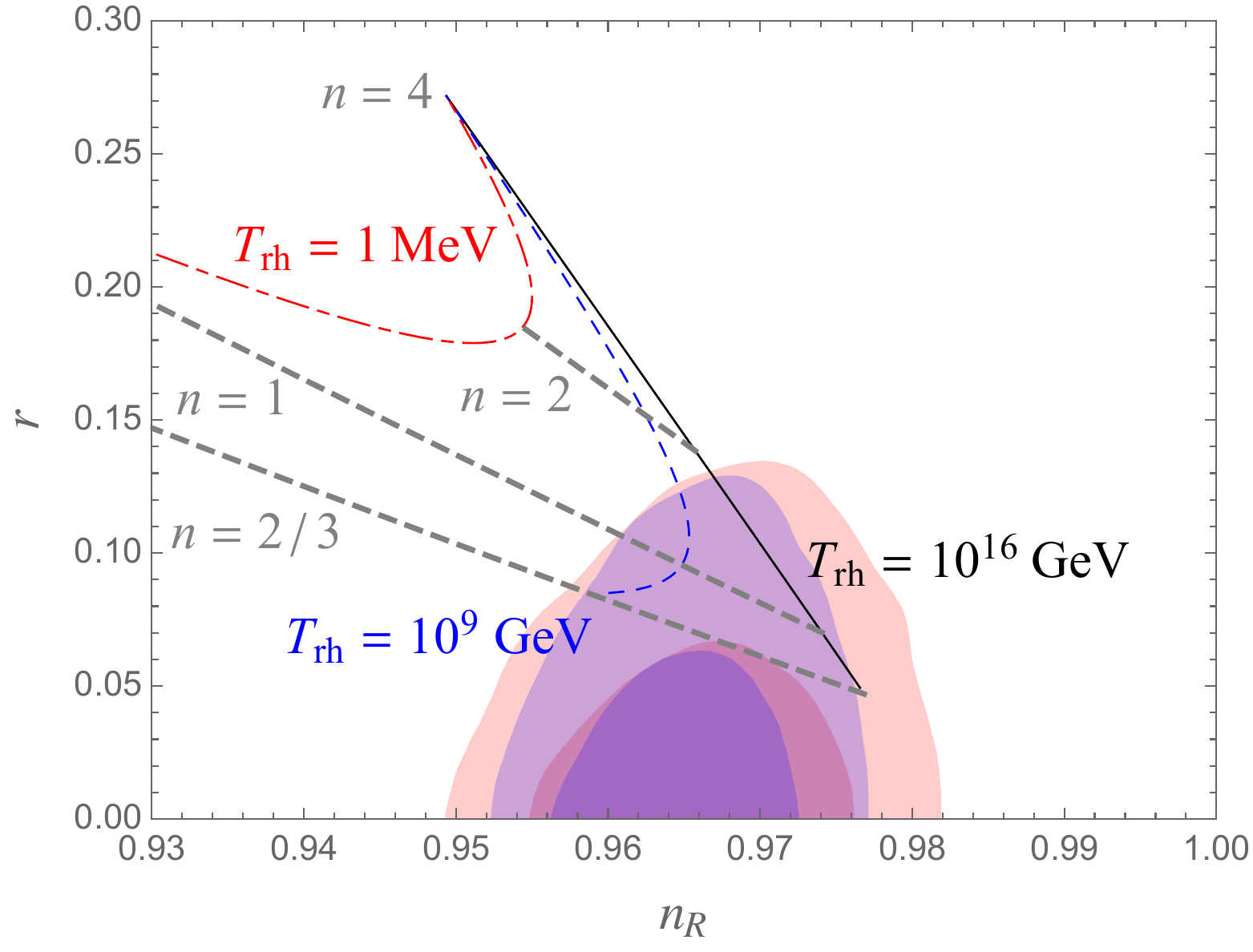}
 \end{center}
 \caption{$n_\calR$-$r$ plot for the power-law chaotic inflation. The rightmost solid line denotes the instantaneous reheating immediately right after inflation ends at around $T_\rh\sim10^{16}~\text{GeV}$. The leftmost dot-dashed line is the lower bound for $T_\rh$ just above the nucleosynthesis bound $1~\text{MeV}$. The intermediate dashed line symbols $T_\rh = 10^9~\text{GeV}$ which is usually taken as the upper bound to avoid gravitino overproduction when supersymmetry is considered. Different power indices for the potential at $n=2/3$, 1, 2 and 4 are marked by thick dashed lines.  The Planck 2015 contours (larger, TT+lowP) and (smaller,TT,TE,EE+lowP) are also shown. Notice that $n=4$ case gives a robust prediction, which, however, is excluded at 99.7\% confidence level by the Planck data.}
 \label{fig:chaotic}
\end{figure}

Let us first we consider simple power-law potential
\begin{equation}\label{power-law}
V(\phi) = \frac{1}{2}m^{4-n}\phi^n \, ,
\end{equation}
with a canonical kinetic term. Its effective equation of state during the oscillating phase can be expressed as~\cite{Turner:1983he}
\begin{equation}\label{eos}
w_\rh \approx \frac{n-2}{n+2} \, ,
\end{equation}
which is nearly constant if $n$ is a global power index. The computation is straightforward even analytically~\cite{monomial,Martin:2014vha,Creminelli:2014fca}, and the model predictions of $n_\calR$ and $r$ with the reheating correction are shown in  Figure~\ref{fig:chaotic}. We can see that lower reheating temperature gives smaller $n_\calR$ and larger $r$, and that the uncertainty vanishes for $V(\phi) \sim \phi^4$. This is because for $n=4$, from \eqref{eos} we have $w_\rh \approx 1/3$ and the last term in (\ref{Nk}) vanishes, thus the oscillation stage is indistinguishable from the radiation dominated epoch and the predictions of $n_\mathcal{R}$ and $r$ become robust. For an exhaustive study on other canonical single field slow-roll inflation models, see e.g.~\cite{Martin:2014vha,Martin:2014nya}.

In fact, \eqref{power-law} belongs to a universality class of models where the spectral index is inversely proportional to $N_k$,
\begin{equation}\label{universal}
n_\calR-1 = -\frac{p}{N_k} \, ,
\end{equation}
where $p=\calO(1)$ is constant and depends on models, e.g. $p=2$ for power-law inflation with $n=2$ in \eqref{power-law}. Besides power-law inflation, this universal scaling relation also applies to a wide range of models~\cite{scaling} including the $R^2$ inflation model~\cite{Starobinsky:1980te}. With this universal scaling relation, we can ask what is the range we can constrain for $N_k$ given a certain accuracy of measurements, which can be translated into that of the reheating temperature $\Delta{T}_\rh$. From \eqref{universal}, the uncertainty in $n_\calR$ can be related to that in $N_k$
\begin{equation}
\label{scaling_ns_deltaNk}
\Delta{N}_k \approx p\frac{\Delta{n}_\calR}{(n_\calR-1)^2} \approx \frac{1-3w_\rh}{12(1+w_\rh)} \log\Delta\beta \, ,
\end{equation}
if we assume the main contribution to $\Delta{N}_k$ comes from the reheating correction term. With an accuracy of $\Delta{n}_\calR = \calO\left(10^{-3}\right)$, which is expected from future observations such as the Euclid satellite~\cite{Amendola:2012ys}, and a fiducial value $n_\calR = 0.968$ from Planck\cite{Ade:2015lrj}, we see this gives $\Delta{N}_k \approx 0.7p \sim 1-2$. Considering a model that gives predictions within the observed range of $n_\calR$, we deduce the uncertainty in $T_\rh$ we can constrain is
\begin{equation}\label{DeltaT}
\frac{\Delta{T}_\rh}{T_\rh} = \calO\left( 10^2-10^3 \right)
\end{equation}
for canonical oscillation around a quadratic minimum, where $w_\rh\approx0$. For non-quadratic potentials the uncertainty can be different, and it can become as small as $\Delta{T}_\rh/T_\rh = \calO(10)$ for $w_\rh \approx -1/3$. On the other extreme case where $w_\rh\approx1/3$, $N_k$  is totally independent of $T_\rh$ and therefore $T_\rh$ is not constrained from $n_\calR$ and $r$. These results are general as long as the model we consider are within the same universality class \eqref{universal}.

We can consider as well a similar universal prediction on the tensor-to-scalar ratio,
\begin{equation}
r = \frac{q}{N_k^m} \, ,
\end{equation}
where $q$ and $m$ are some constants of $\calO(1)$. For instance, quadratic power-law potential gives $q=8$ and $m=1$. The corresponding constraint on $N_k$ by the measurement on $r$ is given by
\begin{equation}
\Delta{N}_k \approx \frac{q^{1/m}}{m} \frac{\Delta{r}}{r^{1+1/m}} \, .
\end{equation}
For practical observability, we may consider only up to $m=2$ that gives $r=\calO(0.01-0.001)$, which is near the sensitivity of future observations on the $B$-mode polarization such as LiteBIRD~\cite{litebird}. For $m=1$, $r$ is typically as large as $\calO(0.1)$. Thus with the prospect $\Delta{r} = 10^{-3}$ we find $\Delta{N}_k = \calO(0.1)$, e.g. for quadratic power-law potential $\Delta{N}_k \approx 0.45$~\cite{Creminelli:2014oaa} so that we can constrain $T_\rh$ with a good accuracy,
\begin{equation}
\frac{\Delta{T}_\rh}{T_\rh} = \calO(1) \, .
\end{equation}
Even for smaller value of $r$ with $m=2$, $\Delta{T}_\rh/T_\rh$ is better than or at least as good as what we can obtain from $n_\calR$: for $r=0.01$, we can find $\Delta{T}_\rh/T_\rh = \calO\left(10-10^2\right)$.

Similarly, we can do the same for the running of the spectral index, $\alpha_\calR$. From \eqref{universal}, in general $\alpha_\calR \propto N_k^{-2}$, so that the constraint on $N_k$ we can obtain from the measurement of $\alpha_\calR$ is
\begin{equation}
\Delta{N}_k \approx s \frac{\Delta\alpha_\calR}{\alpha_\calR^{3/2}} \, ,
\end{equation}
with $s=\calO(1)$ being constant. Given the fiducial value $\alpha_\calR = 10^{-3}$ and the accuracy $\Delta\alpha_\calR = \calO\left(10^{-4}-10^{-5}\right)$ that could be achieved in joint constraints using CMB and future large scale structure or 21cm observations like Square Kilometre Array~\cite{Adshead:2010mc}, we can expect $\Delta{T}_\rh/T_\rh = \calO(10)$. Thus a detection of $\alpha_\calR$ in future experiments would give a better or compatible constraint on $T_\rh$ compared to that given by the measurements of $n_\calR$.

We should stress that in general the predictions of $n_\calR$, $r$ and $\alpha$ do not only depend on the inflation and reheating dynamics, but also on other phases after the end of inflation. Phases additional to the usual post-inflationary cosmic history where reheating is followed by radiation and matter dominated eras, if any, also induce uncertainties in $N_k$ and thus can bias our predictions of the bound of the reheating temperature. We confine ourselves within the minimal scenario and do not consider such scenarios here.

\section{A case study: non-minimal inflation}
\label{sec:nonminimal}

The power-law potential we have discussed in the previous section, which belongs to the universal class \eqref{universal}, is usually regarded as an effective description at around the energy scale of inflation. The effective equation of state \eqref{eos} is only valid when the power index $n$ remains the same both in the large field region and the neighborhood of the minimum. It is therefore interesting to study the extension to this monomial power-law inflation, like for instance the polynomial inflation stemmed from renormalizability~\cite{polynomial1} or some superpotentials~\cite{polynomial2}. These examples can be still described by \eqref{universal} and after we simplify the polynomial to a piecewise potential, all the discussions are parallel. As in~\cite{polynomial1}, a natural cutoff of the power series will be at $n=4$, which means the potential behaves as quartic in the large field region. It is already clear that the constraint on the reheating temperature becomes loose when $n$ is approaching $4$, or, from another point of view, $n\approx4$ gives a robust prediction for $n_\calR$ and $r$ independent of the detail of reheating. However, as now the large-field monomial quartic potential is excluded at $99.7\%$ confidence level by the Planck data~\cite{Ade:2015lrj}, it is worthwhile to seek for other models that are also independent of the reheating temperature $T_\rh$. A natural consideration is to introduce a non-minimal coupling to gravity for the quartic monomial or polynomial up to quartic, which is also physically well motivated by the Higgs inflation model~\cite{Bezrukov:2007ep}. The action in the Jordan frame is
\begin{equation}\label{non-minimal}
S_J = \int d^4x \sqrt{-g} \left[ \frac{\mpl^2}{2} \left( 1+\xi\frac{\phi^2}{\mpl^2} \right) R - \frac{1}{2}g^{\mu\nu}\partial_\mu\phi\partial_\nu\phi - \frac{\lambda}{4}\phi^4 \right] \, .
\end{equation}
In Higgs inflation $\phi$ is identified as the standard model Higgs boson and there is a vacuum expectation value $v \approx 246~\text{GeV}$ in the potential corresponding to the mass $m_H \approx 125~\text{GeV}$~\cite{higgs}, which is neglected here. Performing the conformal transformation of the metric $g_{\mu\nu} \rightarrow \left( 1 + \xi\phi^2/\mpl^2 \right) g_{\mu\nu}$, we can work in the Einstein frame with the following action
\begin{equation}
S_E = \int d^4x\sqrt{-g} \left[ \frac{\mpl^2}{2}R - \frac{1}{2} g^{\mu\nu}\partial_\mu\chi\partial_\nu\chi - W(\chi) \right] \, ,
\end{equation}
where the canonically normalized field in the Einstein frame $\chi$ is connected to $\phi$ by~\cite{non-minimal}
\begin{equation}
\frac{{\rm d}\chi}{{\rm d}\phi} = \frac{\sqrt{1+(1+6\xi)\xi\phi^2/\mpl^2}}{1+\xi\phi^2/\mpl^2} \, ,
\end{equation}
and the effective potential in the Einstein frame $W$ is related to the Jordan frame potential by
\begin{equation}
W(\phi) = \frac{\lambda}{4} \frac{\phi^4}{\left(1+\xi\phi^2/\mpl^2\right)^2} \, .
\end{equation}
Note that since the conformal transformation factor to the Einstein frame is a function of $\phi$ only, perturbations and their correlation functions are equivalent in both frames~\cite{frameequiv} and calculations can be performed in the Einstein frame. For example, the power spectrum of the curvature perturbation is
\begin{equation}
\calP_\calR \approx \frac{\lambda}{768\pi^2} \frac{1+\xi(1+6\xi)\phi_k^2/\mpl^2}{\left( 1+\xi\phi_k^2/\mpl^2 \right)^2} \frac{\phi_k^4}{\mpl^4} \, ,
\end{equation}
which can be used to replace $\lambda$, and the energy density at the end of inflation as
\begin{equation}\label{rhoe}
\frac{\rho_e}{\mpl^4} \approx \frac{3\left( \sqrt{1+64\xi/3+128\xi^2}-1 \right)^2}{8\xi^2 \left( 1+12\xi+\sqrt{1+64\xi/3+128\xi^2} \right)^2}\lambda \, .
\end{equation}

In the minimal scenario of reheating (see also~\cite{Martin:2014vha}), $w_\rh$ depends upon only the field dynamics near the minimum of the potential. For $\xi\ll1$, there is little deviation from the simple quartic power-law potential, so that $w_\text{rh}\approx1/3$ and the predictions for $n_\calR$ and $r$ are fixed irrespective of $T_\rh$. They lie, however, on a curve rather than a single point on the $n_\calR$-$r$ plane depending on the value of $\xi$. On the other hand, for $\xi\gg1$, there exists a critical value $\chi_\text{cr}$ where the form of the potential changes~\cite{Bezrukov:2008ut}:
\begin{equation}\label{Wasymp}
W(\chi) \approx \left\{
\begin{array}{lll}
\dfrac{\lambda}{6\xi^2}\mpl^2\chi^2 & \text{for} & \chi_\text{cr} < \chi \ll \chi_e
\\
\dfrac{\lambda}{4}\chi^4 & \text{for} & \chi < \chi_\text{cr}
\end{array}
\right. \, ,
\end{equation}
where $\chi_e\sim\mpl$ corresponds to the end of inflation and $\chi_\text{cr} \approx \sqrt{2/3}\mpl/\xi$ is the critical point where the dominant term in the potential changes. Thus, right after inflation the oscillation is sinusoidal as in a quadratic potential and $w_\rh \approx 0$. If the inflaton decays before the oscillating amplitude becomes smaller than $\chi_\text{cr}$, reheating happens during this effectively matter dominated stage. If the decay happens later, the potential is quartic and $w_\rh \approx 1/3$. Again, it goes into the region where the prediction on $n_\calR$ and $r$ is independent of the reheating temperature.

We have to find the range of $\xi$ where the approximation \eqref{Wasymp} is valid. We can see from \eqref{Wasymp} that if $\xi$ is not large enough, the energy density at the critical value $\chi_\text{cr}$ will be larger than that at the end of inflation and the potential exhibits only monimial quartic power, leaving no quadratic behavior as in \eqref{Wasymp}. The value of $\xi$ where this happens, $\xi_\star$, can be found by solving $\rho(\chi_\text{cr}(\xi_\star)) = \rho_e(\xi_\star)$ which, with \eqref{rhoe} and \eqref{Wasymp}, gives
\begin{equation}
\xi_\star = \frac23 + \frac{\sqrt{10}}{4} \approx 1.457 \, .
\end{equation}
It is not much larger than 1 but we can still assume that the potential take the form of \eqref{Wasymp} when $\xi>\xi_\star$.

\begin{figure}[h]
 \begin{center}
  \includegraphics[width=0.65\textwidth]{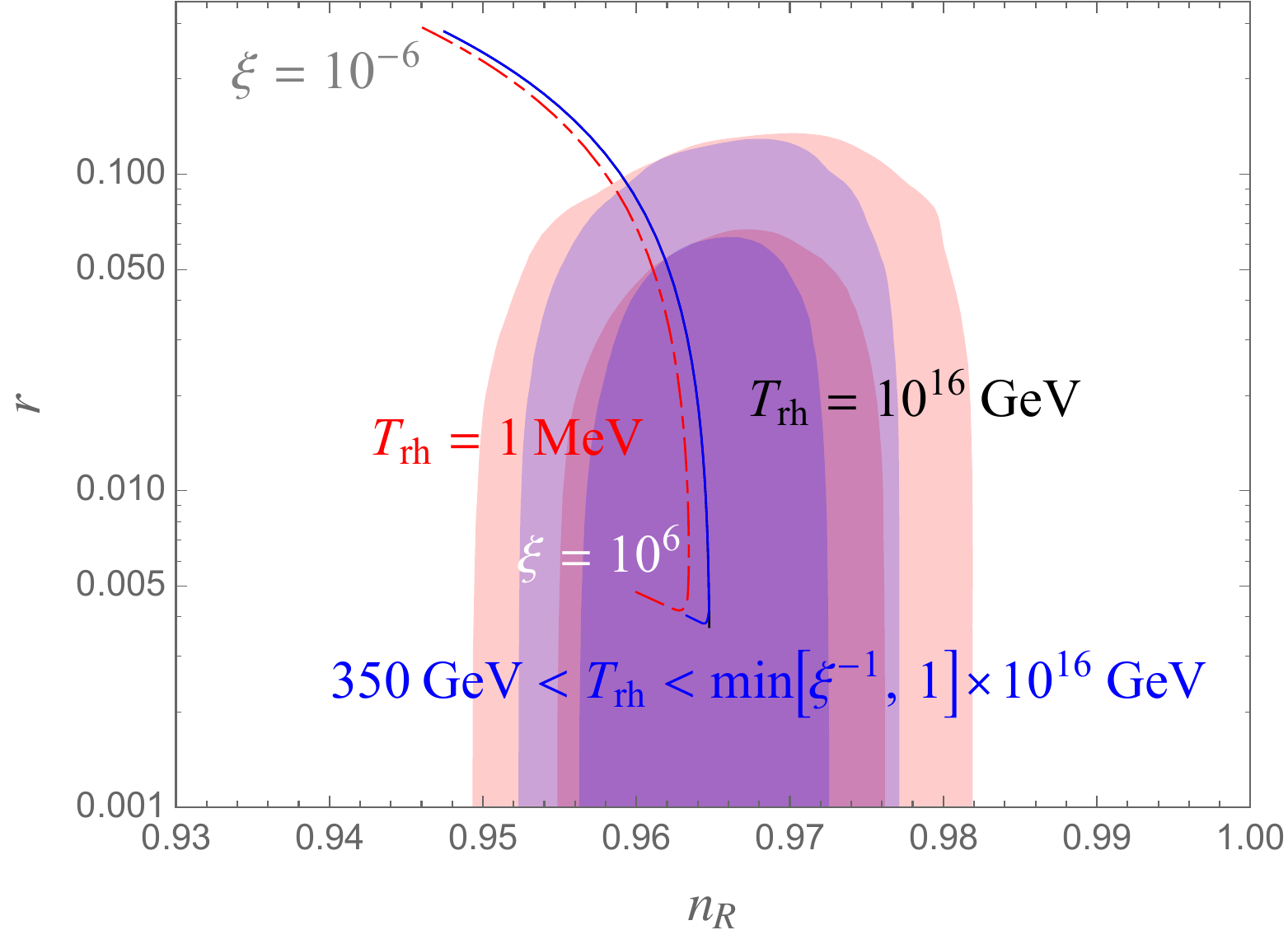}
 \end{center}
 \caption{The predictions on the $n_\calR$-$r$ plane of the non-minimal inflation model \eqref{non-minimal}. The rightmost dashed line corresponds to the instantaneous reheating with $T_\rh \sim 10^{16}~\text{GeV}$, while the solid line represents a huge range of the reheating temperature, $T_\rh\lesssim\min[\xi^{-1},1]\times10^{16}~\text{GeV}$.  This case is almost indistinguishable from the instantaneous reheating for $\xi\lesssim1$, but approaches the predictions of the $R^2$ model for large $\xi$. If we consider a non-zero vacuum expectation value $v$, which we take $v \approx 246~\text{GeV}$, the predictions begin to change when the quadratic oscillation around $v$ becomes effective, corresponding to $T_\rh \approx 350~\text{GeV}$. The leftmost dot-dashed line corresponds to the reheating temperature of $1~\text{MeV}$, with a non-vanishing vacuum expectation value taken into account. The non-minimal coupling runs from $10^{-6}$ (top) to $10^6$ (bottom). All the predictions asymptotes those of the $R^2$ model when $\xi$ is large, but the same point represents different $e$-folding numbers and different reheating temperatures in these two models.}
 \label{fig:nonminimal}
\end{figure}

Using the same procedure as in Section~\ref{sec:general}, we can find $N_k$ similar to \eqref{Nk}. If $\xi<\xi_\star$, the predictions are identical to the instantaneous reheating, for which $N_k$ is given by just \eqref{Nk} with $\beta=1$. If $\xi>\xi_\star$, the intermediate quadratic stage $\chi_\text{cr}<\chi<\chi_e$ also affects the evolution and we have an extended version of \eqref{Nk}:
\begin{align}
\label{Nrh}
N_k & = -\log\left( \frac{k}{a_0H_0} \right) + \log\left( \frac{T_0}{H_0} \right) +  \log\left( \frac{H_k}{\rho_e^{1/4}} \right) + \log \left[ \left( \frac{11}{43}g_{s*} \right)^{-1/3} \left( \frac{\pi^2}{30}g_* \right)^{1/4} \right]
\nonumber\\
& \quad +\left[ \frac{1}{3\left(1+w_1\right)} - \frac{1}{3\left(1+w_2\right)} \right] \log \left( \frac{\rho_{\text{cr}}}{\rho_e} \right) + \frac{1-3w_2}{12(1+w_2)}\log\beta \, ,
\end{align}
where $\rho_\text{cr}$ is the energy density at the critical point $\chi_\text{cr}$, and $w_1$ and $w_2$ denote respectively the effective equations of state for the oscillating phase before and after the critical point $\chi_\text{cr}$. Note that $\rho_\rh$ disappears in the final result as $w_2=1/3$, which means that all the predictions for a reheating temperature smaller than $\rho_\rh^{1/4}$ are the same. This is a strong assert implying the robustness of the predictions for the entire model as we can see in Figure~\ref{fig:nonminimal}. The relation \eqref{Nrh} can be extended to multiple stages, which can be used to study a piecewise potential for example including the final quadratic oscillation around the minimum at $v\approx246~\text{GeV}$ in the Higgs inflation, which is also shown in Figure~\ref{fig:nonminimal}. In general, a polynomial potential, if the characteristic scales for each term to dominate have a hierarchy structure as in \eqref{Wasymp}, can always be approximated by a piecewise potential and studied similarly as in \eqref{Nrh}.

As we can see in Figure~\ref{fig:nonminimal}, the predictions on $n_\calR$-$r$ plane are quite insensitive to $T_\rh$ and are mainly dependent on the value of $\xi$: small $\xi$ corresponds to the simple quartic power-law potential, while large $\xi$ asymptotes the $R^2$ inflation model. However, the same point in the asymptotic line predicted by the non-minimal coupling model with a large $\xi$ and by the $R^2$ inflation model corresponds to different reheating temperatures in these two different models. For example, the point in Figure~\ref{fig:nonminimal} of $(n_\calR,r) = (0.9633, 0.004024)$ represents a huge range of reheating temperature of $350~\text{GeV}\lesssim T_\text{rh}\lesssim 10^{10}~\text{GeV}$ in the non-minimal coupling model, while it only corresponds to a single but much smaller reheating temperature of $1~\text{GeV}$ in the $R^2$ inflation model. It is becasue the effective equation of state $w_\rh$ is different in the two models. In the non-minimal coupling model, $w_\rh\sim 1/3$ over a huge range of reheating temperature whereas $w_\rh\sim 0$ in $R^2$ inflation. Thus although the non-minimal model provides the $1/N$ scaling to the tilt as $R^2$ inflation in the regime we consider, the bounds on $T_\rh$ are much worse here as we can see from \eqref{scaling_ns_deltaNk}. 

The uncertainty from reheating occupies only a very small region in the $n_\calR$-$r$ plane when the potential is approximately quadratic. Note that the reheating temperature for the Higgs inflation lies in the region~\cite{Bezrukov:2008ut}
\begin{equation}
3.4 \times 10^{13}~\text{GeV} < T_\rh < \left( \frac{\lambda}{0.25} \right)^{1/4} \times 1.1 \times 10^{14}~\text{GeV} \, .
\end{equation}
The largest possible uncertainty for $e$-foldings is given by
\begin{equation}
\Delta{N}_k \approx \left| \frac{1}{12}\log\left( \frac{\rho_\text{cr}}{\rho_e} \right) \right| < \frac{1}{6}\log\xi \sim 2.28 \, ,
\end{equation}
for a typical value of $\xi=10^6$. Also, the corresponding $n_\calR$ and $r$ with this uncertainty $\Delta{N}_k$ taken into account are
\begin{equation}
\begin{split}
n_\calR & = 0.9640 \pm 0.0007 \, ,
\\
r & = 0.0040 \pm 0.0003 \, .
\end{split}
\end{equation}
An interesting point is that a lower bound of the tensor-to-scalar ratio $r>0.003714$ is predicted irrespective of $\xi$. Note that a detailed study of reheating including the preheating stage in the Higgs inflation model is presented in~\cite{GarciaBellido:2008ab} in parallel with~\cite{Bezrukov:2008ut}.

\section{Conclusions}
\label{sec:conc}

To conclude, we have studied the contributions of reheating to the predictions of $n_\calR$ and $r$. Assuming perturbative and fast reheating, the uncertainty from reheating is determined by the reheating temperature $T_\rh$ and the effective equation of state during the oscillating stage after inflation, which is determined by the inflation model. We have considered a universal class of predictions on the uncertainty in the reheating temperature, and found that future observations can pin down $T_\rh$ within a factor of $\calO(1)$ optimistically if the oscillation around the minima of inflation potential is sinusoidal and $w_\rh\approx 0$. We have also considered the inflation model with a scalar field non-minimally coupled to gravity, motivated by the Higgs inflation, as a typical example that has a different oscillation behavior. We have found that its predictions are insensitive to the reheating temperature, allowing precise predictions even without any knowledge of reheating. In studying this model, we have established an equation of $N_k$ for models involving different phases of oscillations with different equations of state, which can be easily extended to other cases like polynomial or piecewise potentials.

\subsection*{Acknowledgements}

We thank Qing-Guo Huang and Misao Sasaki for useful discussions. We would also like to thanks the referee for his/her comments.
SP thanks Kavli Institute for Theoretical Physics China for the hospitality when this work was initiated.
We acknowledge the Max-Planck-Gesellschaft, the Korea Ministry of Education, Science and Technology, Gyeongsangbuk-Do and Pohang City for the support of the Independent Junior Research Group at the Asia Pacific Center for Theoretical Physics. This work was also supported in part by a Starting Grant through the Basic Science Research Program of the National Research Foundation of Korea (2013R1A1A1006701).

\end{document}